\documentstyle[prl,aps,epsf,multicol]{revtex} 
\tighten
\renewcommand{\narrowtext} 
{\begin{multicols}{2}\global\columnwidth20.5pc} 
\renewcommand{\widetext}
{\end{multicols}\global\columnwidth42.5pc} 
\begin{document} 
\draft 
\title{Non-adiabatic scattering of a classical particle in an
inhomogeneous magnetic field} 
\author{F. Evers, A. D. Mirlin,$^*$ D. G. Polyakov,$^\dagger$
P. W\"olfle} 
\address{Institut f\"ur Theorie der Kondensierten Materie,
Universit\"at Karlsruhe, 76128 Karlsruhe, Germany} 
\maketitle

\begin{abstract} We study the violation of the adiabaticity of the
electron dynamics in a slowly varying magnetic field. We formulate and
solve exactly a non-adiabatic scattering problem. In particular, we
consider scattering on a magnetic field inhomogeneity which models
scatterers in the composite-fermion theory of the half-filled Landau
level. The calculated non-adiabatic shift of the guiding center is
exponentially small and exhibits an oscillatory behavior related to
the ``self-commensurability" of the drifting cyclotron orbit. The
analytical results are complemented with a numerical simulation.
\end{abstract} 
\narrowtext
 
The dynamics of particles moving in a spatially random magnetic field
(RMF) ${\bf B}({\bf r})$ has been a subject of considerable interest
in the last few years. The particular appeal of the subject is due to
its relevance to a number of models of strongly interacting disordered
electron systems in two dimensions ($2d$). One of the most prominent
examples is the composite-fermion description \cite{halperin93} of a
half-filled Landau level. Within this approach, the electron liquid in
a strong magnetic field is mapped -- by means of a Chern-Simons gauge
transformation -- to a fermion gas subject to a weak effective
magnetic field. Precisely at half-filling, the expectation value of
the Chern-Simons gauge field compensates the effect of the external
magnetic field. The RMF appears in this model after taking static
disorder into account: fluctuations of the local filling factor due to
screening of the random scalar potential yield a local mismatch
between the gauge and external magnetic fields. A number of
observations \cite{willett97} of Fermi-surface features near
half-filling give a strong experimental support to the model of the
effective magnetic field. Apart from the composite-particle models
involving fictituous fields, $2d$ electron systems with a real RMF can
be directly realized in semiconductor heterostructures by attaching to
the latter superconducting \cite{geim92} or ferromagnetic
\cite{ye,zielinski98} overlayers.

The peculiarity of transport properties of $2d$ electrons in a random
field $B({\bf r})$ shows up most distinctly in systems with {\it
smooth} inhomogeneities. The case of long-range disorder is most
important also experimentally -- since the compressible state in a
half-filled Landau level is observed in high-mobility samples. In the
latter, a large correlation radius of potential fluctuations, $d$, is
fixed by a wide ``spacer" between the electron gas and the doped layer
containing ionized impurities. Likewise, inhomogeneities of the RMF
created by the ferromagnetic overlayers in \cite{ye,zielinski98} appear
to be fairly long-ranged. In the composite-fermion model of the
half-filled Landau level, the large value of $d$ (as compared to the
interelectron distance) allows to ignore, in the first instance, the
quantum interference of scattered waves and describe the electron
kinetics as purely classical. Quantum localization effects become
crucial either on exponentially long scales or at strong deviations
from half-filling, thus motivating us to consider a {\it classical}
particle subject to the RMF.

This paper is closely related to the previous study \cite{mirlin98} of
the electron kinetics in the RMF. The conductivity in the field
$B({\bf r})={\overline B}+\delta B({\bf r})$ in the limit of strong
homogeneous component $\overline B$ was shown in \cite{mirlin98} to be
completely determined by weak {\it non-adiabatic} processes in
scattering on the inhomogeneities $\delta B({\bf r})$. A key
ingredient of this transport theory is the non-adiabatic dynamics on
the microscopic scale. The purpose of the present paper is to study
the microscopic processes of the non-adiabatic scattering in detail by
complementing the analytical arguments with results of a numerical
simulation.

To this end we formulate a {\it single}-scattering problem: we
introduce a weak homogeneous gradient of the background magnetic field
and consider the interaction of an electron with an ``impurity"
modeled by a spatially localized perturbation $\delta B({\bf r})$. Far
away from the impurity the electron motion is a slow van Alfv\'en
drift along straight lines -- which are contours of constant magnetic
field -- accompanied by rapid cyclotron gyrations. We consider the
adiabatic limit, where the shift of the guiding center during one
cyclotron period, $\delta$, is much smaller than the size $d$ of the
impurity, $d/\delta\gg 1$. In this case, the particle continues to
drift along the lines of constant $B({\bf r})$, which is a
manifestation of adiabatic invariance. A weak violation of the
adiabatic invariance leads to a small shift, after the scattering
process has taken place, of the magnetic field contour along which the
particle continues its drift. It is this shift, $\Delta\rho$, that
will be studied below, both analytically and numerically. The
corresponding effect in a system with many impurities governs the
magnetoconductivity, as discussed in \cite{mirlin98}.

In terms of the complex coordinate $z=x+iy$, the equation of motion at
the energy $mv_F^2/2$ in our scattering problem reads $\ddot
z=i\Omega(z)\dot z$, where \begin{equation} \Omega
(z)=\omega_c\left[1+\epsilon {y\over R_c}+W(z)\right]~, \end{equation}
$\omega_c=e{\overline B}/mc$, $\overline B$ is the background field at
$y=0$, $R_c=v_F/\omega_c$, the slope $\epsilon$ yields a finite
velocity of the incident particle in $x$ direction, and $W(z)$ is a
``scattering potential". We assume that $\epsilon\ll 1$. The guiding
center coordinate $y$ averaged over the cyclotron orbit, $\rho=\left<
y\right>_c$, plays the role of an impact parameter. The particle
entering the system at $x=-\infty$ with the initial condition
$\lim_{x\to -\infty}\left<y\right>_c=\rho_i$ will leave it at
$x=\infty$ along the trajectory with
$\left<y\right>_c=\rho_i+\Delta\rho$, where $\Delta\rho$ is the
desired non-adiabatic shift. In order to analyze the scattering
problem analytically we assume the impurity field to be weak, $W\ll
1$. Expanding the coordinate $z$ in powers of $W$: $z=z_0+z_1+
\ldots$, we concentrate on the first-order term $z_1$.

The solution $y_0$ in the absence of the impurity can be written in an
implicit form for the initial conditions $z(0)=0$, $\dot z(0)=iv_F$ as
$t(y_0)=\omega_c^{-1}\int_0^{y_0/R_c}dYD^{-1/2}(Y)$, where
$D(Y)=1-Y^2(1+{\epsilon\over 2}Y)^2$. The coordinate $y_0$ is confined
to the region $y_-\leq y_0\leq y_+$, where $y_{\pm}/R_c=(\sqrt{1\pm
2\epsilon}-1)/\epsilon$, and is periodic with the period
$2|t_+-t_-|\equiv 2\pi/\omega$, where $t_\pm=t(y_\pm)$. The velocity
in $x$ direction, $\dot x_0=-\omega_c(y_0+\epsilon y_0^2/2R_c)$, may
be integrated to give $x_0=\left<\dot x_0\right>_ct+\xi$, where
$\xi(t)$ is also periodic with the frequency $\omega$. The leading
terms in the small-$\epsilon$ expansion are: $\left<\dot
x_0\right>_c={\epsilon\over 2}v_F$, \begin{eqnarray}\xi = R_c[(\cos
\omega t-1)-\epsilon(\sin\omega t+{1\over 4}\sin 2\omega t) +
O(\epsilon^2)]~,\\ y_0=R_c[\sin\omega t -{3\over 4}\epsilon
+\epsilon(\cos \omega t -{1\over 4}\cos 2\omega t) +
O(\epsilon^2)]~,\end{eqnarray} and $\omega=\omega_c+O(\epsilon^2)$. To
find $y_1$, we integrate the equation of motion for $x_1$ once and use
the relation $\dot x_1\dot x_0+\dot y_1 \dot y_0 =0$ following from
energy conservation, which yields a first-order differential equation
for $y_1$ with periodic coefficients \begin{equation} \dot y_0\dot
y_1-\ddot y_0 y_1=\omega_c\dot x_0\int_0^tdt'W[z_0(t')]\dot y_0(t')~.
\end{equation} For each half-period between the turning points
$t=t_\pm$, at which $\dot y_0$ changes sign, the solution of Eq.\ (4)
can be represented in the form \begin{eqnarray} y_1(t)=\omega_c\dot
y_0(t)\int_0^t dt'K(t')\int_0^{t'} dt''W[z_0(t'')]\dot y_0(t'')~.
\end{eqnarray} Here the factor $K(t)=\dot x_0(t)/\dot y_0^2(t)$
behaves singularly as $(t-t_\pm)^{-2}$ in the vicinity of
$t_\pm$. Matching two branches of $y_1(t)$ across the turning point
requires that the contour of the $t'$ integration in Eq.\ (5) be
displaced into the complex plane so as to pass round the singularity
on the real axis [since the term $(t'-t_\pm)^{-1}$ is absent in the
integrand, the contour can be shifted in either half-plane]. This
corresponds to the contraction
$\int^{t_\pm+\tau}_{t_\pm-\tau}dt(t-t_\pm)^{-2}\to -2\tau^{-1}$ at
$\tau\to 0$. With this choice of the contour of integration, Eq.\ (5)
gives the sought solution at all $t$.

To extract the non-adiabatic shift from Eq.\ (5), we observe that the
integral over $t''$ converges to a constant at large $t'$ while
$\left<\dot y_0\right>_c=0$, so that the average
$\Delta\rho=\left<y_1(t\to \infty)\right>_c-\left<y_1(t\to
-\infty)\right>_c$ is given by \begin{equation} \Delta\rho=\alpha
I~,\quad I=\int_{-\infty}^{\infty}dt W[z_0(t)]\dot y_0(t)~,
\end{equation} where $\alpha(\epsilon)=\omega_c\left<\dot
y_0(t)\int_0^t dt'K(t')\right>_c$ is expressed in terms of the
unperturbed solution. At $\epsilon\to 0$ the constant $\alpha\to
-1$. Note that, apart from the shift $\Delta\rho$, the asymptotics of
Eq.\ (5) at large $t$ contains an oscillating term $-\epsilon
I\omega_c t\cos\omega_c t$, whose amplitude diverges linearly with
growing $t$. This divergency is an artefact of the perturbation
expansion in $W$ and reflects the fact that the shift $\Delta\rho$ is
accompanied by the change of the frequency
$(\epsilon\omega_c/R_c)\Delta\rho$. This asymptotics could have been
equivalently used to find $\Delta\rho=-I$.

In evaluating $I$ we first assume, for simplicity, that $W(z)$ depends
on $x$ only, which allows to expand $I$ as \begin{equation}
I=\int_{-\infty}^\infty dt\sum_n {1\over n!}
\partial^n_xW\left({\epsilon\over 2}v_Ft\right)\xi^n(t)\dot y_0(t)~.
\end{equation} Since $W({\epsilon\over 2}v_Ft)$ is a smooth function
of $t$ on the scale of $\omega^{-1}$, the leading contribution to $I$
comes from taking the first harmonic of the integrand -- higher
harmonics will involve exponentially smaller Fourier components of
$W({\epsilon\over 2}v_Ft)$. This yields, after integration by parts
and resummation, \begin{eqnarray} &&I\simeq 2v_F{\rm
Re}\left[A(\epsilon)\int_{-\infty}^\infty dt
e^{i\omega_ct}W\left({\epsilon\over 2}v_Ft\right)\right]~,\\
&&A(\epsilon)={1\over R_c}\int^{2\pi\over\omega_c}_0{dt\over
2\pi}\exp\left(- {2i\xi(t)\over\epsilon R_c}-i\omega_ct\right)\dot
y_0(t)~.\end{eqnarray} The integral in Eq.\ (9) can be evaluated at
$\epsilon\ll 1$ by the saddle-point method to give \begin{eqnarray}
\Delta\rho&=&-2v_F\sqrt{\epsilon\over\pi}\cos\left({2\over\epsilon}-
{\pi\over 4}\right)\\ \nonumber &\times&\int_{-\infty}^\infty dt \cos
\omega_ct\,\,W\left({\epsilon\over 2}v_Ft-R_c\right)~. \end{eqnarray}
This equation is a parametrically exact solution of the scattering
problem at $\epsilon\to 0$. It expresses the non-adiabatic shift in
terms of the asymptotics of the Fourier transform of the smooth
function $W$ -- thus demonstrating explicitly the exponential
smallness of $\Delta\rho$. The parameter which governs the exponential
falloff of $\Delta\rho$ is $d/\delta\gg 1$, where $d$ is a
characteristic scale of variation of $W$ and $\delta=\pi\epsilon R_c$
is the shift of the guiding center after one period of the cyclotron
rotation, while the ratio $d/R_c$ may be arbitrary. Though the limit
of a smooth inhomogeneity on the scale of $R_c$ is historically most
closely associated with the notion of adiabaticity, the parameter
$d/R_c$ plays no role in Eq.\ (10). It is worth noting that the
pre-exponential factor given by the first line of Eq.\ (10) is a
non-analytic function of the small parameter $\epsilon$ and, moreover,
happens to oscillate wildly as $\epsilon\to 0$. These oscillations are
due to the commensurability of two length scales $R_c$ and
$\delta$. Remarkably, the series of the geometric resonances which
constitute the oscillations is associated with the properties of the
unperturbed solution (``self-commensurability") and not with the shape
of the scatterer. Note that to get the oscillations one has to sum up
all terms in the expansion (7) even if $W$ is smooth on the scale of
$R_c$. Another peculiar feature of the non-adiabatic shift is related
to its sensitivity to the phase $\phi$ of the cyclotron rotation of
the incident electron. Specifically, changing the position of the
scatterer by $\Delta x$ leads to the oscillations of $\Delta\rho
(\phi)=\Delta\rho_m\cos(\phi-\phi_0)\propto \int_{-\infty}^\infty
dt\cos(\omega_ct+\phi)W({\epsilon\over 2}v_Ft)$, where
$\phi=2\pi\Delta x/\delta$, so that the shift vanishes periodically
with varying initial conditions.

In the composite-fermion problem, a charged impurity located the
distance $d$ from the plane occupied by the electron gas creates the
axially symmetric perturbation $W({\bf r})=W_0d^3[({\bf r}-{\bf
R})^2+d^2]^{-3/2}$. We choose the impurity position ${\bf R}=
(0,\rho_i)$, so that $\rho_i$ has the meaning of the impact parameter
with which the guiding center is incident on the impurity. Extracting
the first-order Fourier component of $W[z_0(t)]$ in the same way as in
Eqs.\ (8),(9), we get \begin{eqnarray}I&\simeq& 2 {\rm Re}
\int_0^{2\pi\over\omega_c} {dt\over 2\pi} e^{-{2i\xi (t)\over\epsilon
R_c}-i\omega_c t}\dot y_0(t)\\ \nonumber
&\times&\omega_c\int_{-\infty}^\infty
dt'e^{i\omega_ct'}W\left[{\epsilon\over 2}v_Ft',y_0(t)\right]~.
\end{eqnarray} The function $W$ in this equation has branch points at
$t'=t'_s(t)$, where $t'_s(t)=\pm {2i\over\epsilon
v_F}\sqrt{d^2+(y_0(t)-\rho_i)^2}$, which determines the exponentially
small value of the integral over $t'$. Specifically, the second line
in Eq.\ (11) reads $\sqrt{2\pi}W_0(2\pi d/\delta)^3(\omega_c
|t'_s(t)|)^{-3/2}\exp (-\omega_c |t'_s(t)|)$. The remaining integral
over $t$ can be done by the saddle-point method. The cumbersome
general expression reduces to \begin{eqnarray} \Delta\rho&\simeq& 8\pi
W_0 (\rho_i+\tilde{d})\left({R_cd\over
\delta^2}\right)^{1/2}e^{-\rho_i/\tilde{d}}\\ \nonumber &\times&
\cos\phi\cos\left({2d\over\epsilon\tilde d}-{\pi\over
4}\right)\exp\left(-2\pi{\tilde{d}\over\delta}\right)~ \end{eqnarray}
in the limit of a long-range impurity potential, when $d,|\rho_i|\gg
R_c$. Here $\tilde{d}=(d^2+\rho_i^2)^{1/2}$. In the opposite case,
when $R_c$ is larger than both $d$ and $|\rho_i|$, we get
\begin{equation} \Delta\rho\simeq 8\pi W_0{d^2\over\delta}\cos\phi\cos
{2\over\epsilon} \exp \left( -2\pi{d\over\delta}\right)~.
\end{equation} The last three factors in Eqs.\ (12),(13) reflect the
features of the non-adiabatic shift discussed above: the exponential
smallness, the oscillations with changing $\epsilon$, and the
oscillatory dependence on the phase $\phi$. Note that all the formulas
above imply that the drift trajectory is only slightly perturbed by
$W({\bf r})$.
 
Now let us turn to a numerical simulation. A typical trajectory
resulting from the numerical integration of Eq.\ (3) is shown in Fig.\
1. Though the pertubation of the drift trajectory is seen to be large
in this particular example, the non-adibatic shift is almost
invisible. The inset to Fig.\ 2 shows the periodic dependence of the
shift on the phase $\phi$, which agrees with Eq.\ (12). The main panel
clearly demonstrates the adiabatic character of the scattering at
$\tilde{d}/\delta\gg 1$: the magnitude of the oscillations
$\Delta\rho_m$ plotted against $\delta^{-1}$ is seen to fall off
exponentially, also as predicted by Eq.\ (12). The slope $c$ of the
exponential decay $\ln\Delta\rho = -2\pi c\tilde{d}/\delta$ is found
to be 1.3, which is somewhat larger than the value $c=1$ following
from Eq.\ (12). This is because the perturbation of the drift
trajectory is not negligible in this case. No oscillations with
changing $\epsilon$ could be reliably seen in Fig.\ 2, which should
also be ascribed to their smearing due to the curvature of the
trajectory. Indeed, the commensurability of $R_c$ and $\delta$ cannot
be maintained in the whole region of interaction if the drift
trajectory is strongly perturbed, as clearly illustrated by Fig.\
1. The non-adiabatic shift at a smaller amplitude of the interaction
$W_0$ is shown in Fig.\ 3. In this case, the perturbation of the
trajectory was weak, and the particle was essentially drifting right
``through the impurity" with a small impact parameter
$\rho_i<d,R_c$. Decreasing $W_0$ narrows the range of the numerical
simulation but clearly reveals the oscillatory behavior of
$\Delta\rho_m$ with changing $\epsilon$.  The exponential dependence
in Fig.\ 3 is fitted very well by $e^{-2\pi d/\delta}$ (in accordance
with Eq.\ (13)) with no fitting parameter in the exponent. The
oscillations of $\Delta\rho_m$ with changing $R_c/\delta$ (inset of
Fig.\ 3) have a period close to $1/2$ and are in good agreement with
the analytical prediction $\Delta\rho_m\propto |\cos {2\pi
R_c\over\delta}|$.

\vspace{-10cm} \begin{figure}[tb]
\centerline{\epsfxsize=72mm\epsfbox{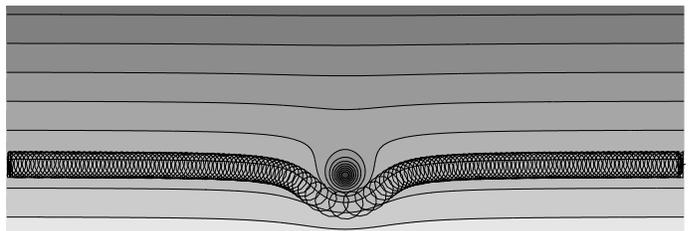}\vspace*{3mm}} \vspace{-7cm}
\caption{Typical trajectory of a particle scattered on a ``magnetic
impurity" [Eq.\ (1)]. The lines of constant magnetic field $B({\bf
r})$ are shown. The strength of the impurity $W_0=2.2$.}
\label{fig1} \end{figure}

\begin{figure}[tb]
\centerline{\epsfxsize=72mm\epsfbox{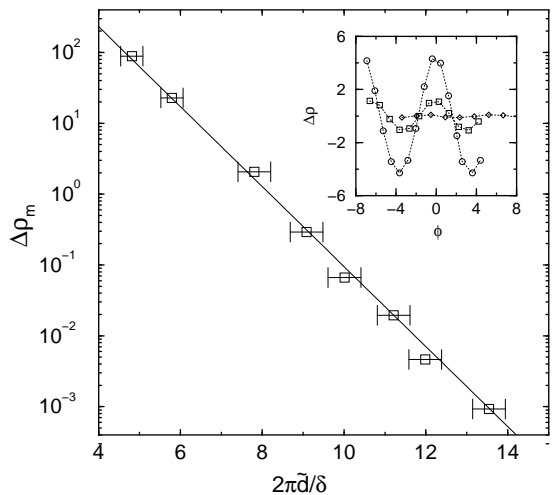}\vspace*{3mm}}
\caption{Amplitude of the non-adiabatic shift $\Delta\rho_m$ as a
function of $\delta=2\pi v_d/\omega$ for a strong impurity:
$W_0=2.2$. The solid line is a fit $\Delta\rho_m\propto\exp
[-1.3\times2\pi \tilde{d}/\delta]$. Inset: Shift $\Delta\rho$ as a
function of the phase $\phi$ for different values of the parameter
$2\pi\tilde{d}/\delta=4.8 (\bigcirc), 5.8 (\Box), 7.8 (\Diamond)$.}
\label{fig2} \end{figure}

\begin{figure}[tb]
\centerline{\epsfxsize=72mm\epsfbox{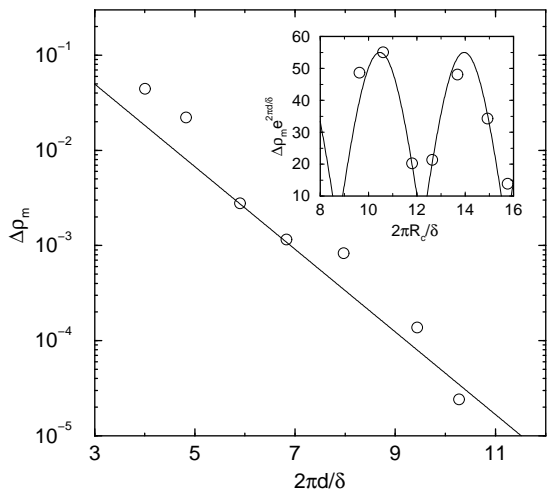}\vspace*{3mm}}
\caption{Non-adiabatic shift for a weak impurity ($W_0= 0.07$) at
$\rho_i/d=0.4-0.9$ and large $R_c/d=1.5-2.5$. The solid line is a fit
$\Delta\rho_m\propto\exp (-2\pi d/\delta)$. Inset: Oscillations of
$\Delta\rho_m$ as a function of $R_c/\delta$.}  \label{fig3}
\end{figure}

In conclusion, we have studied the violation of the adiabaticity of
the electron dynamics in a slowly varying magnetic field. We
formulated a scattering problem which has been solved exactly. As a
particular example we considered scattering on a single ``magnetic
impurity" which models a scatterer in the composite-fermion theory of
the half-filled Landau level. The non-adiabatic shift of the guiding
center is exponentially small and exhibits oscillations with
$R_c/\delta$, where $R_c$ is the cyclotron radius, $\delta$ the shift
of the guiding center along the drift trajectory after one cyclotron
period. The oscillations are related to the ``self-commensurability"
of the drifting cyclotron orbit. The analytical results are in full
agreement with the numerical simulation. A detailed numerical study of
the transport properties of the fermions in a random magnetic field
will be presented elsewhere.

This work was supported by the Deutsche Forschungsgemeinschaft through
SFB 195 and by the Graduiertenkolleg ``Kollektive Ph\"anomene im
Festk\"orper''.

\end{multicols}
\end{document}